\documentclass[%
 reprint,
 superscriptaddress,
 bibnotes,
 amsmath,amssymb,
 aps,
]{revtex4-1}

\usepackage{graphicx}
\usepackage{subfigure}
\usepackage{dcolumn}
\usepackage{bm}
\usepackage[colorlinks, linkcolor=blue, urlcolor=blue, citecolor=blue]{hyperref}
\usepackage{float}
\usepackage[utf8]{inputenc}
\usepackage{epstopdf}
\begin{document}

\preprint{APS/123-QED}

\title{Electrically controllable thermal transport in Josephson junctions based on buckled two-dimensional materials}

\author{Yu-Hao Zhuo}
%\email{hnnuhl@hunnu.edu.cn}
\affiliation{Key Laboratory of Low-Dimensional Quantum Structures and Quantum Control of Ministry of Education, Key Laboratory for Matter Microstructure and Function of Hunan Province, School of Physics and Electronics, Hunan Normal University, Changsha 410081, China}

\author{Biao Wu}
%\email{hnnuhl@hunnu.edu.cn}
\affiliation{Key Laboratory of Low-Dimensional Quantum Structures and Quantum Control of Ministry of Education, Key Laboratory for Matter Microstructure and Function of Hunan Province, School of Physics and Electronics, Hunan Normal University, Changsha 410081, China}

\author{Gang Ouyang}
%\email{gangouy@hunnu.edu.cn}
\affiliation{Key Laboratory of Low-Dimensional Quantum Structures and Quantum Control of Ministry of Education, Key Laboratory for Matter Microstructure and Function of Hunan Province, School of Physics and Electronics, Hunan Normal University, Changsha 410081, China}

\author{Hai Li}
\email{hnnuhl@hunnu.edu.cn}
\affiliation{Key Laboratory of Low-Dimensional Quantum Structures and Quantum Control of Ministry of Education, Key Laboratory for Matter Microstructure and Function of Hunan Province, School of Physics and Electronics, Hunan Normal University, Changsha 410081, China}

%\date{\today}

\begin{abstract}
 We investigate the thermal transport properties in superconductor- antiferromagnet- superconductor and superconductor- ferromagnet- superconductor junctions based on buckled two-dimensional materials (BTDMs). Owing to the unique buckled sublattice structures of BTDMs, in both junctions the phase dependence of the thermal conductance can be effectively controlled by perpendicular electric fields. The underlying mechanism for the electrical tunability of thermal conductance is elucidated resorting to the band structures of the magnetic regions. We also reveal the distinct manifestations of antiferromagnetic and ferromagnetic exchange fields in the thermal conductance. These results demonstrate that the perpendicular electric field can serve as a knob to externally manipulate the phase-coherent thermal transport in BTDMs-based Josephson junctions.
\end{abstract}

\maketitle

\section{\label{sec:level1} Introduction}

  The thermal transport in temperature-biased Josephson junctions has recently garnered considerable attention, due to the extensive applications ranging from phase-coherent caloritronics \cite{Pekola2021, Giazotto2012, Fornieri2016, Timossi2018, Fornieri2017b, Fornieri2017a, Shelly2016, Bauer2021, Bours2019, Mukhopadhyay2022, Hajiloo2019, Pershoguba2019, Sothmann2017} to the detection of novel quantum states \cite{Sothmann2016, Bauer2021L, Mukhopadhyay2021, Gresta2021, Bours2018, Blasi2020, Li2017, Wang2022, Bauer2019}. In Josephson junctions, the formation of Andreev-bound states (ABSs) has a profound impact on the quasiparticle scattering. Since the binding energy and spectral weight of the ABSs depend on the superconducting phase difference, the coupling between the quasiparticles and ABSs gives rise to a phase-coherent component of the thermal current \cite{Kazumi1965, Guttman1997, Zhao2003, Zhao2004}. This effect holds the promise to manipulate the thermal transport via the phase coherence intrinsic to superconducting condensates, boosting the efforts to design phase-coherent caloritronics devices based on Josephson junctions \cite{Pekola2021, Giazotto2012, Fornieri2016, Timossi2018, Fornieri2017a, Fornieri2017b}. During the last decade, there has been tremendous experimental progress in the realm of phase-coherent caloritronics, such as the realization of heat interferometer \cite{Giazotto2012}, heat modulator \cite{Fornieri2016}, thermal router \cite{Timossi2018}, and thermal tunable $0-\pi$ phase transition \cite{Fornieri2017b} in temperature-biased Josephson junctions.

  \par
  On the other hand, since the thermal currents are mainly carried by quasiparticles with energies above the superconducting gap, they provide complementary information to the charge currents which derive essentially from the ABSs and quasiparticles with energies below the superconducting gap \cite{Kazumi1965, Guttman1997, Zhao2003, Zhao2004, Ren2013, Sothmann2016}. In this regard, the thermal transport measurement opens a compensate route to identify the existence of novel quantum states. Recent theoretical proposals have shown that the thermal currents in temperature-biased topological Josephson junctions can be used to probe the topological ABSs \cite{Sothmann2016}, Majorana zero modes \cite{Bauer2021L, Mukhopadhyay2021}, Jackiw-Rebbi resonant states \cite{Gresta2021}, and helical edge states \cite{Bours2018, Blasi2020}. Furthermore, since the thermal transport is sensitive to the pairing symmetry of the superconducting condensate, the thermal transport signature can sever as a valuable tool to distinguish the spin-singlet and spin-triplet pairing states in temperature-biased topological \cite{Li2017} and conventional Josephson junctions \cite{Bauer2019}.

  \par
  Although significant achievements have been made in the thermal transport properties of Josephson junctions, the research attention to date has mainly been restricted to the phase-coherent aspect of thermal transport \cite{Giazotto2012, Fornieri2016, Timossi2018, Fornieri2017a, Fornieri2017b, Shelly2016, Bauer2021, Bours2019, Mukhopadhyay2022, Sothmann2016, Sothmann2017, Bauer2021L, Mukhopadhyay2021, Gresta2021, Bours2018, Blasi2020, Bauer2019, Li2017, Kazumi1965, Guttman1997, Zhao2003, Zhao2004, Wang2022}. In practice, the manipulation of the proposed phase dependence needs to resort to an external magnetic field \cite{Giazotto2012, Fornieri2016, Timossi2018, Fornieri2017a, Fornieri2017b}. It is natural to ask that whether the phase-coherent thermal transport can be managed in a fully electric manner. An exciting possibility is to consider the thermal transport in Josephson junctions based on buckled two-dimensional materials (BTDMs) with electrically tunable low-lying physics.

  \par
  BTDMs are atomically thin crystals possessing hexagonal lattice structures and Dirac-like low-energy excitations, commonly known as silicene, germanene and stanene \cite{Molle2017, Zheng2020, Kezerashvili2021, Zhao2016, Chen2016, Gori2019, Chiappe2014, Grazianetti2018, Feng2019, Wiggers2019, Jabra2022, Liu2011, Drummond2012}. Since a stable BTDM sheet prefers a buckled sublattice structure, the low-energy bands and relevant transport properties can be effectively modulated by an electric field perpendicular to the sheet plane \cite{Liu2011, Drummond2012, Tsai2013, Chen2018, Niu2019, Lu2020, Rojas-Brise2021, Yokoyama2013}. Moreover, recent efforts have predicted that the superconducting correlations can be induced in BTDMs through the proximity effect \cite{Moun2022, Ezawa2015, Wei2021, Linder2014}. This progress together with the unique buckled geometry render BTDMs fertile playgrounds to explore the electrically tunable phase-coherent transport properties \cite{Frombach2018, Kuzmanovski2016, Li2016a, Zhou2017, Paul2017, Li2016b, Lu2021, Paul2016}. One of the most prominent examples is the occurrence of electrically controlled $0-\pi$ phase transition in silicene-based Josephson junctions \cite{Kuzmanovski2016, Li2016a, Zhou2017}. Additionally, recent advances have also revealed that both the local and nonlocal Andreev reflections in silicene-based superconducting hybrid structures can be regulated by a perpendicular electric field \cite{Linder2014, Paul2017, Li2016b, Lu2021}. However, up to now the effects of perpendicular electric field on the phase-coherent thermal transport have been scarcely studied in BTDMs-based Josephson junctions.

  \par
  Motivated by the significance but the lack of detailed understanding about the electrically tunable thermal transport in BTDMs-based Josephson junctions, in this work we investigate the thermal transport properties in superconductor-antiferromagnet-superconductor (S-AF-S) and superconductor-ferromagnet-superconductor (S-F-S) junctions based on BTDMs. Since the perpendicular electric field can modulate the band structures of BTDMs, the phase dependence of thermal conductance is electrically controllable in both S-AF-S and S-F-S junctions. Taking advantage of the band structures in the magnetic regions, the electrical tunability of phase-coherent thermal conductance is elucidated. We also illustrate the different manifestations of antiferromagnetic and ferromagnetic exchange fields in the thermal conductance. Our findings suggest that the perpendicular electric field can be employed to externally manipulate the phase-coherent thermal transport in BTDMs-based Josephson junctions.
  \par
  The rest of this paper is organized as follows. We present the model and calculation method in Sec.~\ref{sec:level2}. In Sec.~\ref{sec:level3}, we give the numerical results and discuss the effects of the perpendicular electric field on the thermal conductance. Finally, the conclusion is briefly drawn in Sec.~\ref{sec:level4}.

\section{\label{sec:level2}  Model and approach}

  \begin{figure}
  \centering
  \includegraphics[width=7cm]{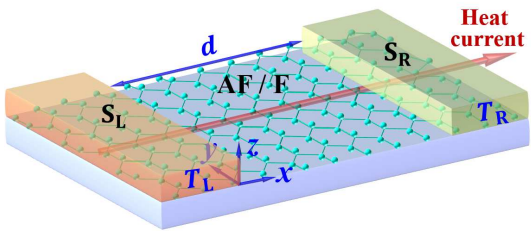}
  %\vspace{2.5cm}
  \caption{(Color online) Sketch of a BTDM-based Josephson junction with a heat current flowing along the $x$-direction.}
  \label{fig:fig1}
  \end{figure}

  \par
  A schematic of the proposed Josephson junction is shown in Fig.~\ref{fig:fig1}, where a BTDM sheet is deposited in the $xy$ plane, with two superconducting electrodes $S_L$ and $S_R$ covering the left (L, $x<0$) and right (R, $x>d$) regions, respectively. To drive a thermal current flowing along the $x$ direction, a temperature gradient is imposed across the junction, where the temperature in the L (R) region is fixed as $T_{L(R)}$ with $T_{L(R)}=T+(-)\delta T/2$. The superconductivity in the L and R regions can be induced by the superconducting electrodes via the proximity effect, as that have been experimentally carried out in similar two-dimensional materials such as graphene \cite{Heersche2007, Bretheau2017, Perconte2018} and transition-metal dichalcogenides \cite{Trainer2020, Dreher2021}. In the magnetic region (M, $0<x<d$) of the S-AF-S (S-F-S) junction, an antiferromagnetic (ferromagnetic) exchange field is introduced to regulate the thermal transport. As demonstrated by recent experiments, the proposed exchange fields can be realized in BTDMs by intercalating rare-earth atoms \cite{Tokmachev2017, Tokmachev2018}.

  \par
  In the superconducting regions, we take the intra-sublattice Bardeen-Cooper-Schrieffer pairing, as proposed in Refs. \cite{Linder2014, Frombach2018, Kuzmanovski2016, Li2016a, Zhou2017}. In the basis of $\psi^\dag_k=\{(\psi_{k, \sigma}^A)^\dag, (\psi_{k, \sigma}^B)^\dag, \psi_{-k, \bar \sigma}^A, \psi_{-k, \bar \sigma}^B\}$ spanned in the Nambu $\otimes$ sublattice space, the Bogoliubov-de Gennes (BdG) Hamiltonian is given by
  \begin{equation}
  {\cal H} = \left( {\begin{array}{*{20}c}
  {H_0-\sigma h} & {\sigma \Delta} \\
  {\sigma \Delta^{\dag}} & {- (H_0- \bar \sigma h)} \\
  \end{array}} \right),
  \label{eq:Hamiltonian}
  \end{equation}
  where the spin index $\sigma=\pm1$ satisfying $\sigma= - \bar \sigma$. The single-particle effective Hamiltonian $H_0=\hbar v_F (\eta k_x \tau_x - k_y \tau_y) + m_{\eta \sigma} \tau_z - \mu \tau_0$ \cite{Liu2011, Drummond2012, Tsai2013, Chen2018, Niu2019, Lu2020, Rojas-Brise2021, Yokoyama2013}, where $\tau_j$ ($j=x, y, z$) denotes the Pauli matrix operating in the sublattice space, $\tau_0$ is a $2 \times 2$ unit matrix, $v_F$ represents the Fermi velocity, and $\eta = +(-)1$ labeling the $K(K^\prime)$ valley. The effective-mass term $m_{\eta \sigma}=lE_z - \eta \sigma \lambda_{\mathrm {SO}}$,  where $\lambda_{\mathrm {SO}}$ indicates the strength of spin-orbit coupling, $E_z$ parameterizes the perpendicular electric field, and $2l$ is the separation between the A and B sublattices along the $z$ direction. In the S-AF-S and S-F-S junctions, the exchange fields are, respectively, characterized as $h=h_{AF}\tau_z \Theta(x) \Theta(d-x) $ and $h=h_F \tau_0 \Theta(x) \Theta(d-x)$, with $\Theta(x)$ the Heaviside step function. The chemical potential $\mu =\mu_S \Theta(-x) + \mu_M \Theta(x) \Theta(d-x) + \mu_S \Theta(x-d)$. In this paper, we take the superconducting regions to be heavily doped to satisfy the relation of $\mu_S \gg \mu_M$, so that the leakage of Cooper pairs from the superconducting regions to the magnetic region can rationally be neglected \cite{Linder2014, Frombach2018, Kuzmanovski2016, Li2016a, Zhou2017, Paul2017, Li2016b, Lu2021, Paul2016}. In doing so, the superconducting gap can be effectively modeled by a step function, i.e., $\Delta = \Delta_{L} (T_L) \tau_0 e^{i \phi_L} \Theta (-x) +  \Delta_{R} (T_R) \tau_0 e^{i \phi_R } \Theta (x-d)$, with the phase difference being defined as $\phi = \phi_R - \phi_L$. The amplitude of the superconducting gap is given by $\Delta_{L(R)} (T_{L(R)}) = \Delta_0 \tanh (1.74 \sqrt{T_C/T_{L(R)}-1})$, where the critical temperature $T_C=\Delta_0/(1.76 k_B)$ and $k_B$ denotes the Boltzmann constant \cite{Bauer2021, Hajiloo2019} .

  \par
  In the present work, we study the thermal transport properties by virtue of the scattering wave approach. This method has been extensively employed to investigate the thermal transport properties in temperature-biased superconducting hybrid structures \cite{Sothmann2016, Sothmann2017, Bours2018, Bours2019, Pershoguba2019, Ren2013, Hajiloo2019, Li2017, Gresta2021, Mukhopadhyay2021, Mukhopadhyay2022, Blasi2020, Bauer2019, Bauer2021L, Wang2022}. Compared with the approaches of tunneling Hamiltonian and Usadel equation in quasiclassical approximation, the scattering wave method possesses the advantages to explore the thermal transport properties in a single- or a few-channel superconducting hybrid structures with arbitrary transparency \cite{Pershoguba2019}.
  \par
  To evaluate the thermal conductance in the proposed Josephson junctions, we first compute the quasiparticle transmission probabilities. For an electron-like (a hole-like) quasiparticle incident from the L region, the resulting wave function $\Psi^{e(h)}_L$ is given by
  \begin{equation}
  \Psi^{e(h)}_L = \psi^{L, +}_{eq(hq)} + r^{ee(hh)}_{\eta\sigma} \psi^{L, -}_{eq(hq)} + r^{he(eh)}_{\eta\sigma} \psi^{L, -}_{hq(eq)},
  \end{equation}
  where $r^{ee, hh}_{\eta\sigma}$ and  $r^{he, eh}_{\eta\sigma}$  denote the valley- and spin-resolved scattering amplitudes of normal reflections and Andreev reflections, respectively. The corresponding wave function in the R region is formulated as
  \begin{equation}
  \Psi^{e(h)}_R = t^{ee(hh)}_{\eta\sigma} \psi^{R, +}_{eq(hq)} + t^{he(eh)}_{\eta\sigma} \psi^{R, +}_{hq(eq)},
  \end{equation}
  with $t^{ee, hh, he, eh}_{\eta\sigma}$ the valley- and spin-resolved transmission amplitudes. The details of electron-like (hole-like) scattering states $\psi^{L, \pm}_{eq(hq)}$ and $\psi^{R, +}_{eq(hq)}$are presented in Appendix \ref{sec:levela}. In the M region, the wave function $\Psi_M$ is a linear superposition of all possible scattering states, i.e.,
  \begin{equation}
  \Psi_M = c_1 \psi^+_e + c_2 \psi^-_e + c_3 \psi^+_h + c_4 \psi^-_h,
  \end{equation}
  where the scattering amplitudes are denoted by $c_j$ ($j=1, 2, 3, 4$) and the detailed structures of scattering states $\psi^\pm_{e, h}$ are given in Appendix \ref{sec:levela}.

  \par
  The transmission amplitudes can be obtained by matching the relevant wave functions at boundaries $x=0$ and $x=d$. To take into account the influences stemming from the interface imperfection, at the boundary $x=0$ $(d)$ we introduce an ultra-narrow square potential barrier characterized by strength $U_{L(R)}$ and width $\ell_{L(R)}$, and then take the limits of $U_{L(R)} \to \infty$ and $\ell_{L(R)} \to 0$ with $U_{L(R)}\ell_{L(R)}/(\hbar v_F) \equiv Z_{L(R)}$ being finite. According to the conservation of the particle current flowing along the $x$ direction, the boundary conditions can be formulated as
  \begin{subequations}
  \begin{equation}
  \Psi^{e(h)}_L|_{x=0^-} = {\cal M}^{-1}_L \Psi_M|_{x=0^+},
  \label{BC1}
  \end{equation}
  \begin{equation}
  \Psi^{e(h)}_R|_{x=d^+} = {\cal M}_R \Psi_M|_{x=d^-},
  \label{BC2}
  \end{equation}
  \label{BC}
  \end{subequations}
  with the transfer matrix ${\cal M}_{L(R)}$ being defined as
  \begin{equation}
  {\cal M}_{L(R)} = e^{i \nu_0 \tau_x \eta Z_{L(R)}},
  \label{eq:sbd}
  \end{equation}
  where $\nu_0$ denotes an unit matrix operating in the Nambu space.

  \par
  Resorting to the transmission amplitudes, the total valley- and spin-resolved transmission probability resulting from the electron-like and hole-like incident quasiparticles can be obtained as
  \begin{widetext}
  \begin{equation}
  {\cal T}_{\eta \sigma}(\epsilon, \theta) = \left|\frac{\langle \psi^{R, +}_{eq} | {\hat j}_x | \psi^{R, +}_{eq} \rangle}{\langle \psi^{L, +}_{eq} | {\hat j}_x | \psi^{L, +}_{eq} \rangle}\right| |t^{ee}_{\eta \sigma}|^2
  + \left|\frac{\langle \psi^{R, +}_{hq} | {\hat j}_x | \psi^{R, +}_{hq} \rangle}{\langle \psi^{L, +}_{eq} | {\hat j}_x | \psi^{L, +}_{eq} \rangle}\right| |t^{he}_{\eta \sigma}|^2
  + \left|\frac{\langle \psi^{R, +}_{eq} | {\hat j}_x | \psi^{R, +}_{eq} \rangle}{\langle \psi^{L, +}_{hq} | {\hat j}_x | \psi^{L, +}_{hq} \rangle}\right| |t^{eh}_{\eta \sigma}|^2
  + \left|\frac{\langle \psi^{R, +}_{hq} | {\hat j}_x | \psi^{R, +}_{hq} \rangle}{\langle \psi^{L, +}_{hq} | {\hat j}_x | \psi^{L, +}_{hq} \rangle}\right| |t^{hh}_{\eta \sigma}|^2,
  \end{equation}
  \end{widetext}
  where the particle current density operator ${\hat j}_x \equiv \frac{-i}{\hbar}[x, H_{\mathrm{BdG}}]= \eta v_F \nu_z \tau_x$, with $\nu_z$ the Pauli matrix operating in the Nambu space.

  \par
  We note that, as proposed in Ref.\cite{Ren2013}, the phonon contribution to the thermal transport can be profoundly suppressed by the interface imperfection between the superconducting and magnetic regions. Therefore, we only concentrate on thermal conductance contributed by electron-like and hole-like quasiparticles and neglect the contribution from phonons. Taking advantage of the transmission probability ${\cal T}_{\eta \sigma}(\epsilon, \theta) $, the heat current can be written as \cite{Sothmann2016, Sothmann2017, Bauer2021L, Bours2018, Bours2019, Pershoguba2019, Ren2013, Hajiloo2019, Li2017, Gresta2021, Mukhopadhyay2021, Mukhopadhyay2022, Blasi2020, Bauer2019, Wang2022}
  \begin{equation}
  J\!= \!\frac{1}{h}\!\sum_{\eta \sigma} \!\int^\infty_{\Delta(T)} \!{d\epsilon \! \int^{\pi/2}_{-\pi/2} \! {\cos \theta d\theta \epsilon {\cal{T}}_{\eta \sigma}(\epsilon, \theta) [f(\epsilon, T_L)\!-\!f(\epsilon, T_R)]}},
  \label{eq:hc}
  \end{equation}
  where the Fermi distribution function $f(\epsilon, T_{L(R)})=[e^{\epsilon/(k_B T_{L(R)})}+1]^{-1}$ and $\Delta (T) = \max (|\Delta_L(T)|, |\Delta_R(T)|) $.

  \par
  For the temperature bias $\delta T \rightarrow 0$, the thermal conductance in the linear response regime can be defined as $\tilde \kappa = (J/\delta T)_{\delta T \rightarrow 0}$, and which can be explicitly formulated as
  \begin{equation}
  {\tilde \kappa} = \frac{1}{h} \sum_{\eta \sigma} {\int^\infty_{\Delta(T)} d\epsilon {\int^{\pi/2}_{-\pi/2} \frac{ {\cal T}_{\eta \sigma}(\epsilon, \theta) \epsilon^2 \cos \theta d\theta}{4 k_B T^2 \cosh^2(\frac{\epsilon}{2k_B T})}}}.
  \label{eq:ThC}
  \end{equation}
  To normalize the thermal conductance, it is convenient to introduce a quantity of $\kappa_0 = 4 \pi^2 k_B^2 T/(3 h)$, where $\pi^2 k_B^2 T/(3 h)$ is the thermal conductance quantum \cite{Schwab2000} and the factor 4 takes the valley and spin indices into account. In doing so, the normalized thermal conductance can be expressed as $\kappa = {\widetilde \kappa} /\kappa_0$.

  \begin{figure*}
  \centering
  \subfigure{
  \includegraphics[width=7cm]{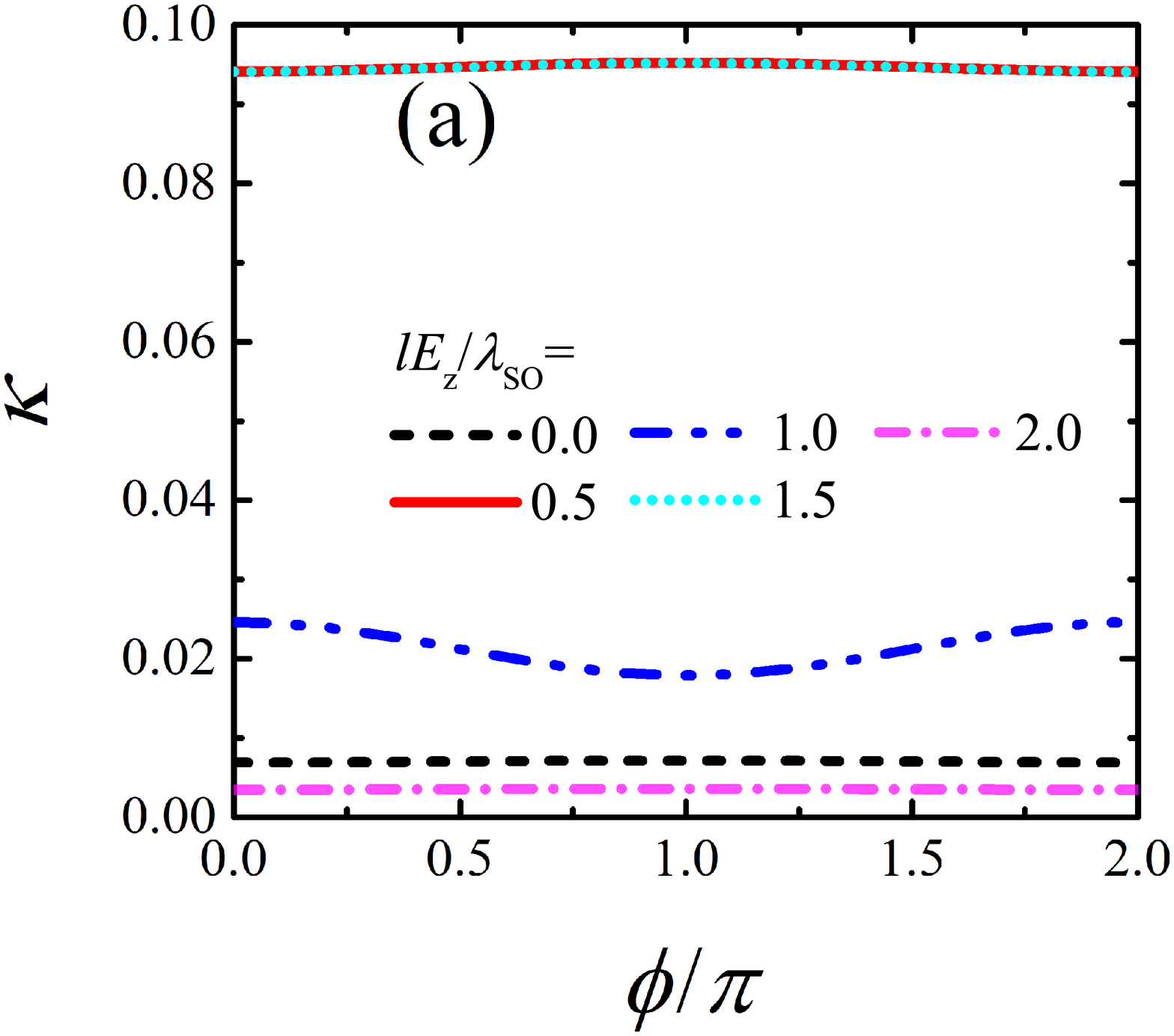}
  \label{subfig:fig2a}}
  \vspace{0.0cm}
  \hspace{2.0cm}
  \subfigure{
  \includegraphics[width=7cm]{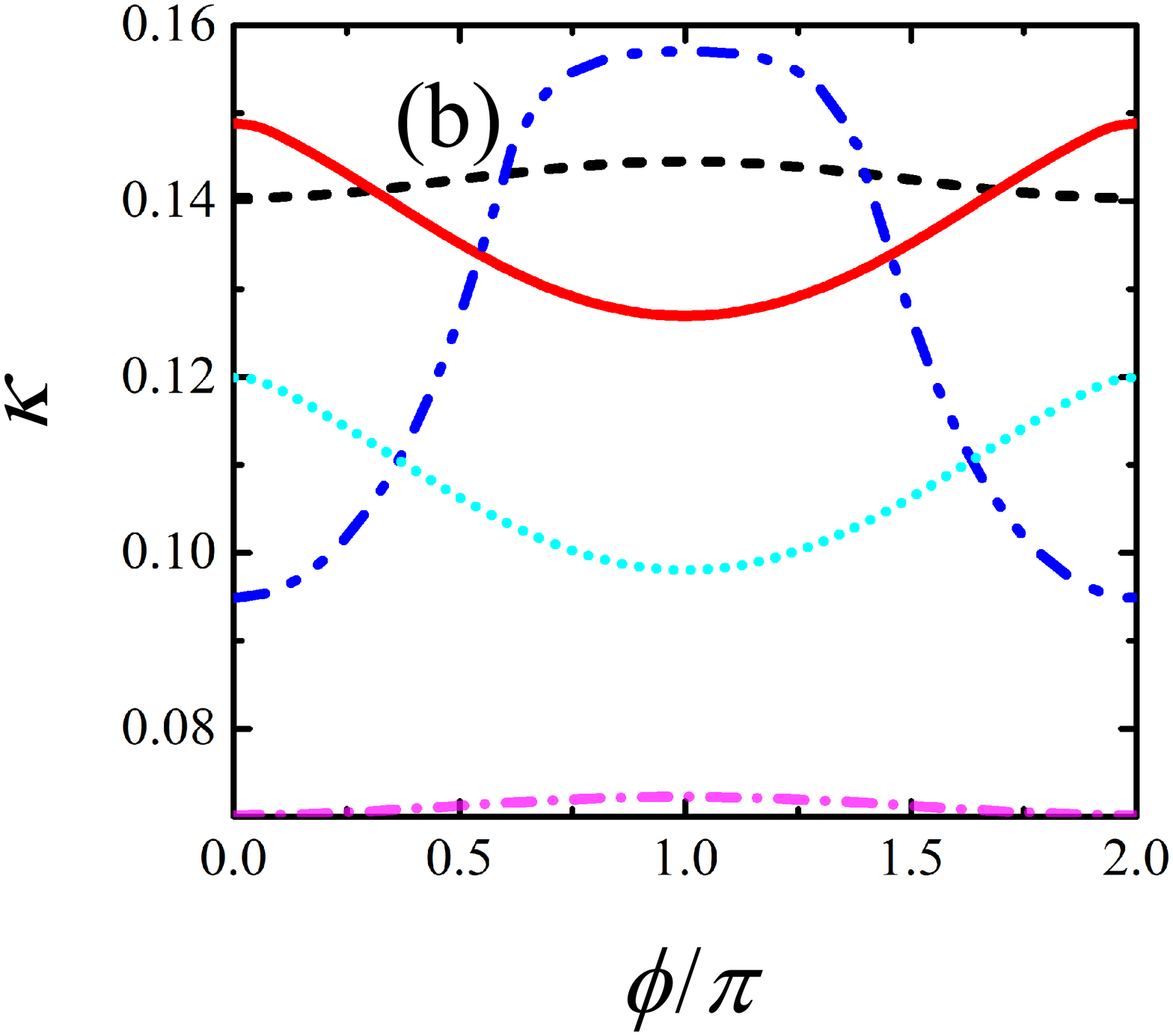}
  \label{subfig:fig2b}}
  \vspace{0.0cm}
  \hspace{0.0cm}
  \subfigure{
  \includegraphics[width=7cm]{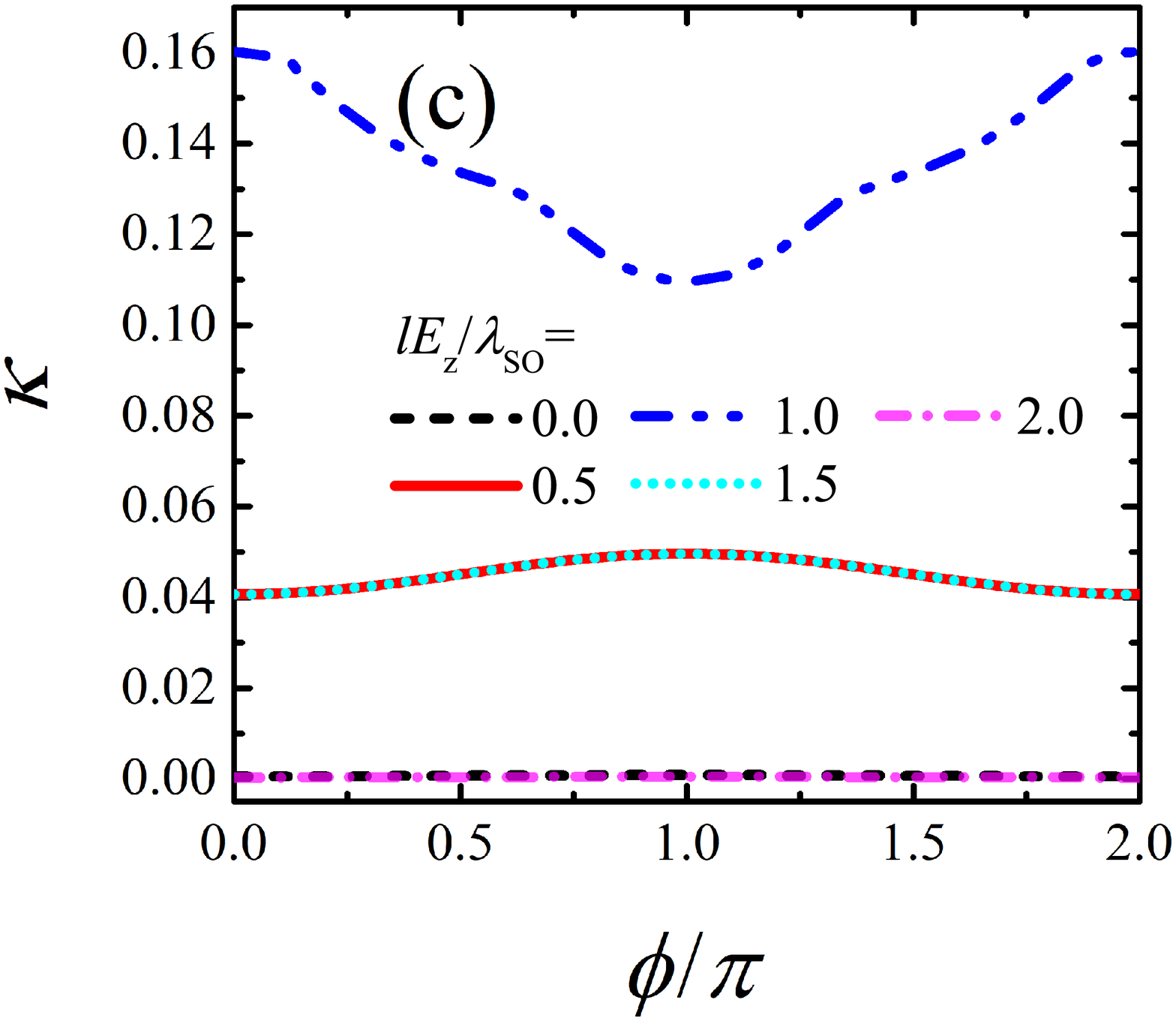}
  \label{subfig:fig2c}}
  \vspace{0.0cm}
  \hspace{2.0cm}
  \subfigure{
  \includegraphics[width=7cm]{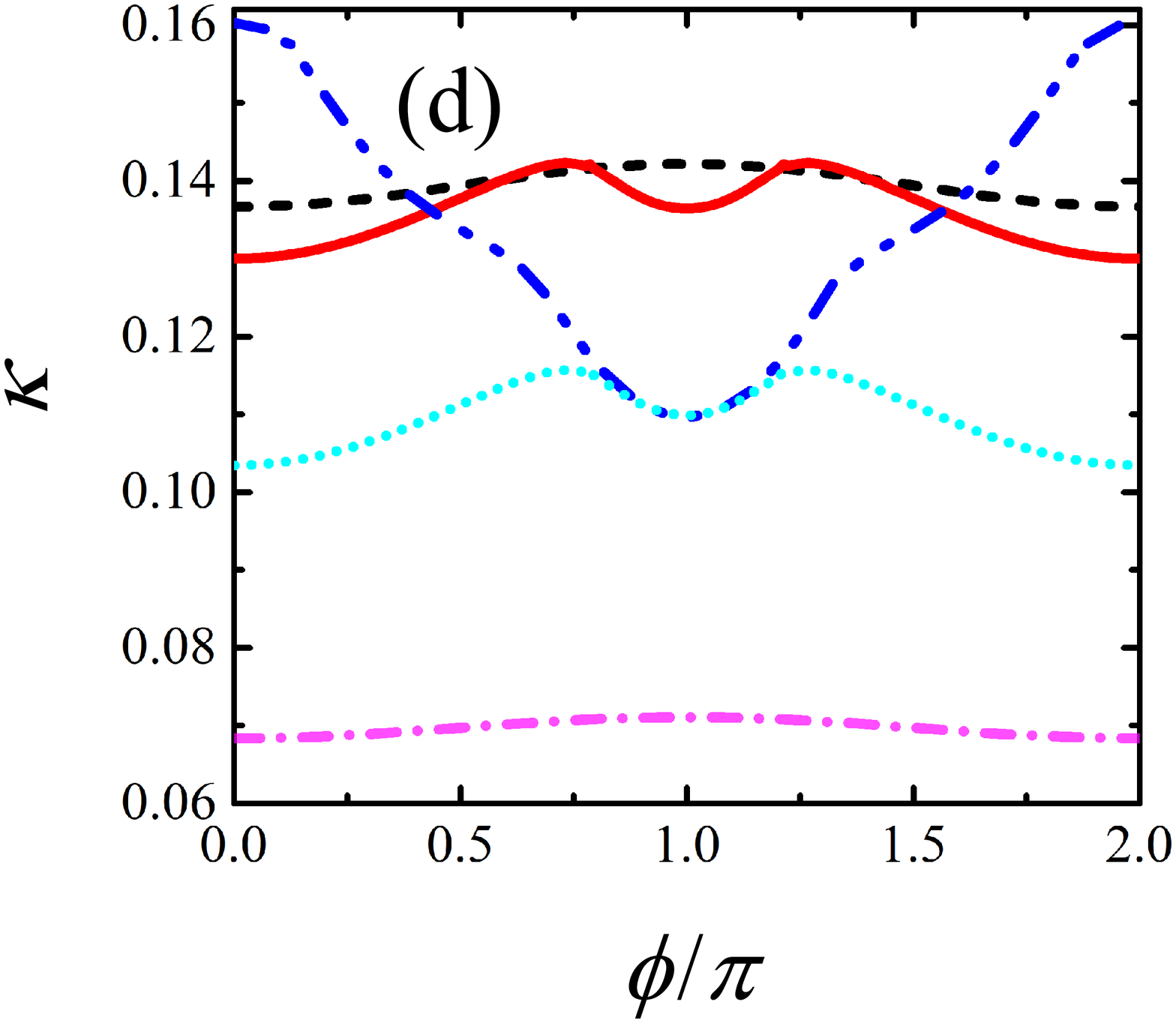}
  \label{subfig:fig2d}}
  \vspace{0.0cm}
  \hspace{0.0cm}
  \caption{(Color online) Phase-difference-dependent normalized thermal conductance with different perpendicular electric fields, where $\mu_M = 0$ in (a) and (c), and $\mu_M = \lambda_\mathrm{SO}$ in (b) and (d). Panels (a) and (b) and panels (c) and (d) present the results in the S-AF-S junction with $h_{AF}=0.4 \lambda_\mathrm{SO}$ and the S-F-S junction with $h_F= 0.4 \lambda_\mathrm{SO}$, respectively. In all panels, $d=\xi$ and $T=T_C/2$. }
  \label{fig:fig2}
  \end{figure*}

\section{\label{sec:level3} Results and Discussion}

  In this section, we present the numerical results and concentrate on the effects of the perpendicular electric field on the thermal conductance. Without loss of generality, we choose silicene as a prototype of BTDM with $\lambda_{\mathrm {SO}}=3.9$ meV \cite{Liu2011}. Considering that the typical value of the superconducting gap magnitude is of the order of $\sim 1$ meV \cite{Linder2014, Frombach2018, Kuzmanovski2016, Li2016a, Zhou2017, Heersche2007, Bretheau2017, Perconte2018}, we take $\Delta_0=0.2 \lambda_{\mathrm {SO}}$ in the numerical calculation. The superconducting coherence length is defined as $\xi=\hbar v_F /\Delta_0$. To ensure the validity of the model described in Eq.~(\ref{eq:Hamiltonian}), we set $\mu_S = 100 \lambda_{\mathrm {SO}}$ to satisfy $\mu_S \gg \Delta(T)$ and retain the relationship of $\mu_S \gg \mu_M$ throughout this work. In doing so, the quasiparticle scattering angles in the superconducting regions turn to $\theta^{L, R}_{eq, hq} \simeq 0$ and the relevant scattering problems reduce into one-dimensional scenarios, as that have been implemented in a series of studies \cite{Sothmann2016, Bauer2019, Pershoguba2019, Hajiloo2019, Sothmann2017}. In this paper, we are not interested in the effects of the interfacial potential barriers on the thermal conductance and single out $Z_L = Z_R = \pi$ for definiteness, since the influences of interfacial potential barriers on the superconducting coherent transport have been intensively investigated \cite{Sothmann2016, Ren2013, Paul2016, Paul2017}. It is well known that the transmission probability and resulting conductance periodically oscillate with respect to $Z_{L, R}$ without decaying profiles, this phenomenon is a typical hallmark of the momentum-spin/pseudospin locking in Dirac materials.

  \begin{figure*}
  \centering
  \subfigure{
  \includegraphics[width=7cm]{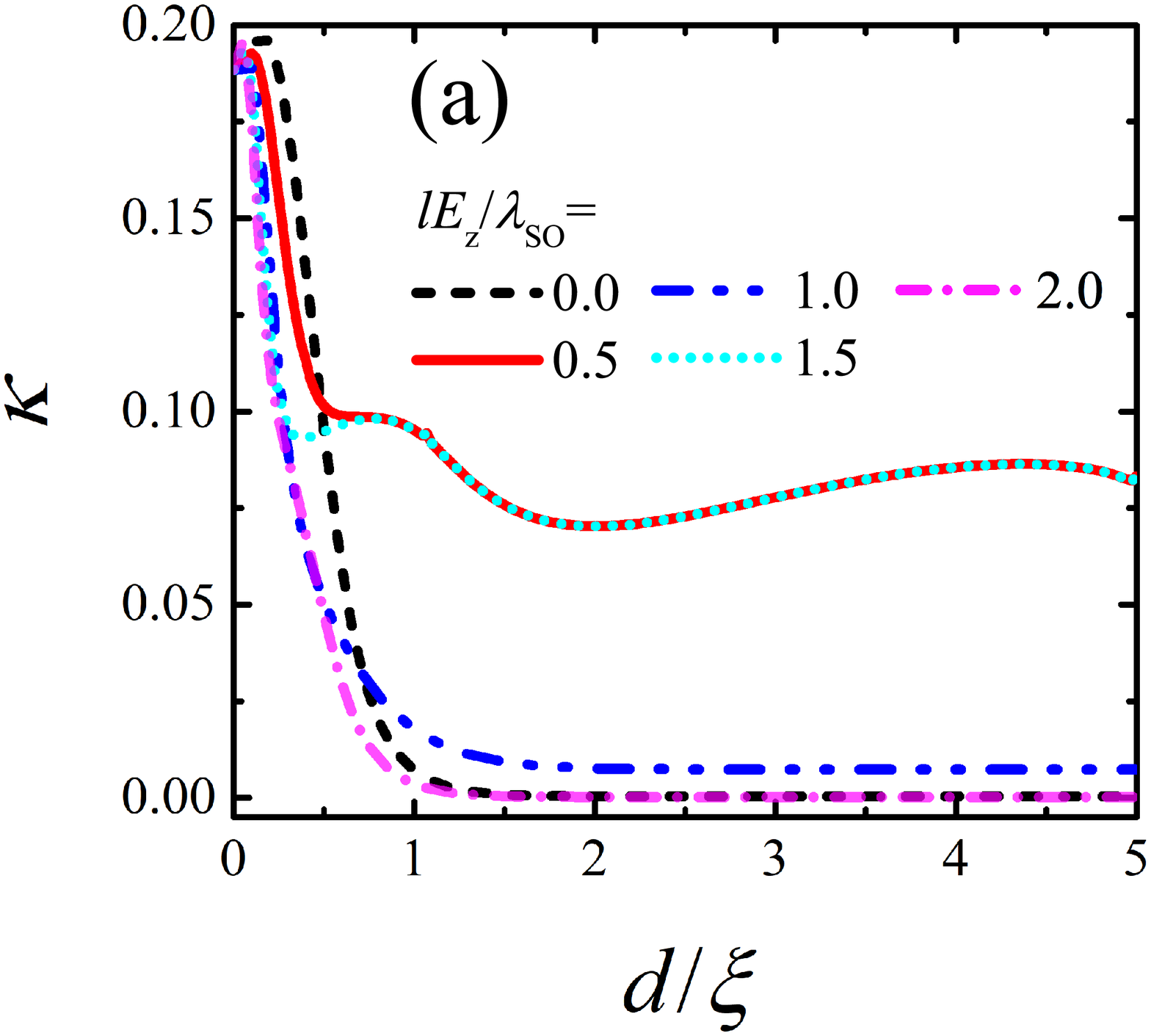}
  \label{subfig:fig3a}}
  \vspace{0.0cm}
  \hspace{2.0cm}
  \subfigure{
  \includegraphics[width=7cm]{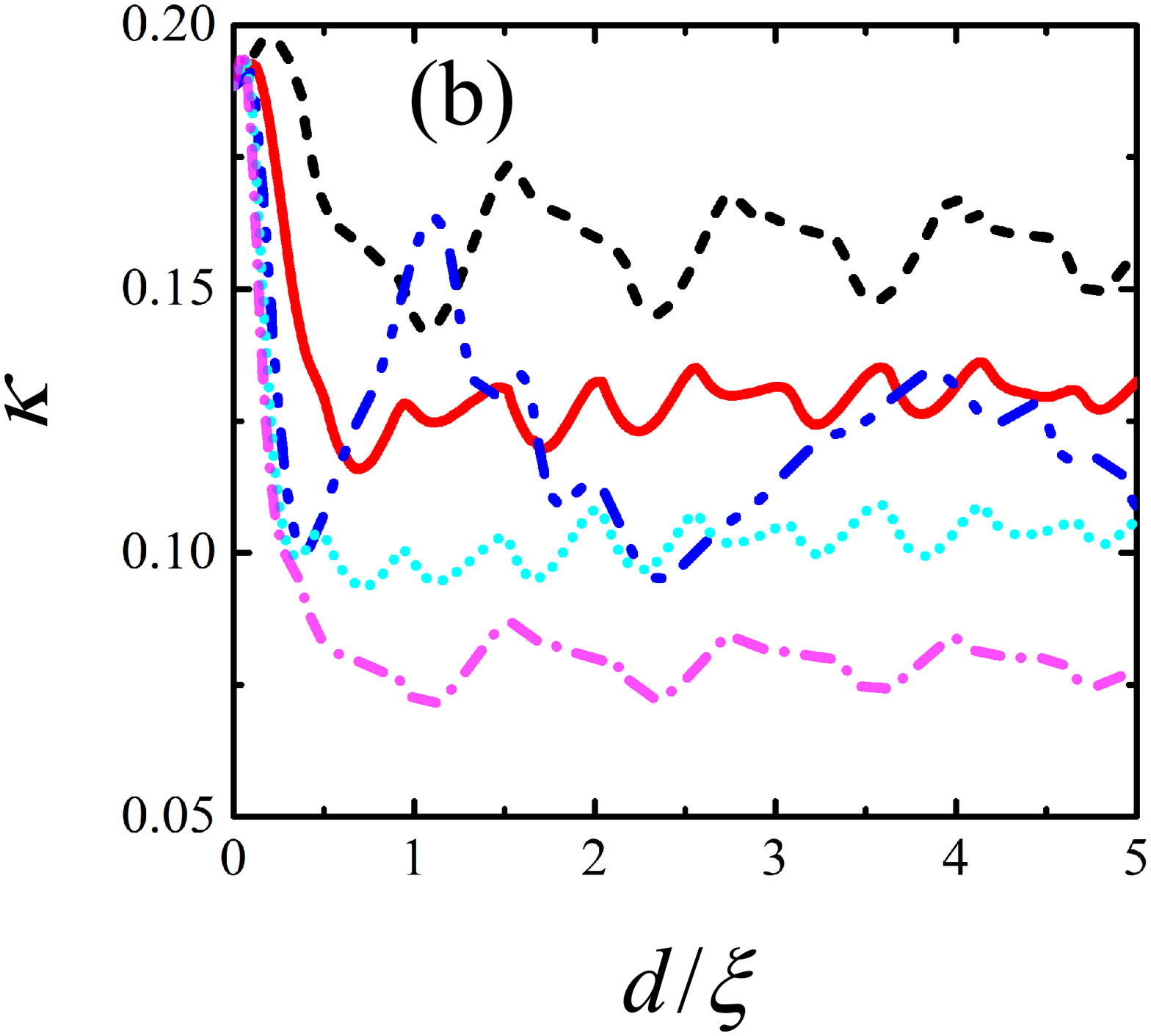}
  \label{subfig:fig3b}}
  \vspace{0.0cm}
  \hspace{0.0cm}
  \subfigure{
  \includegraphics[width=7cm]{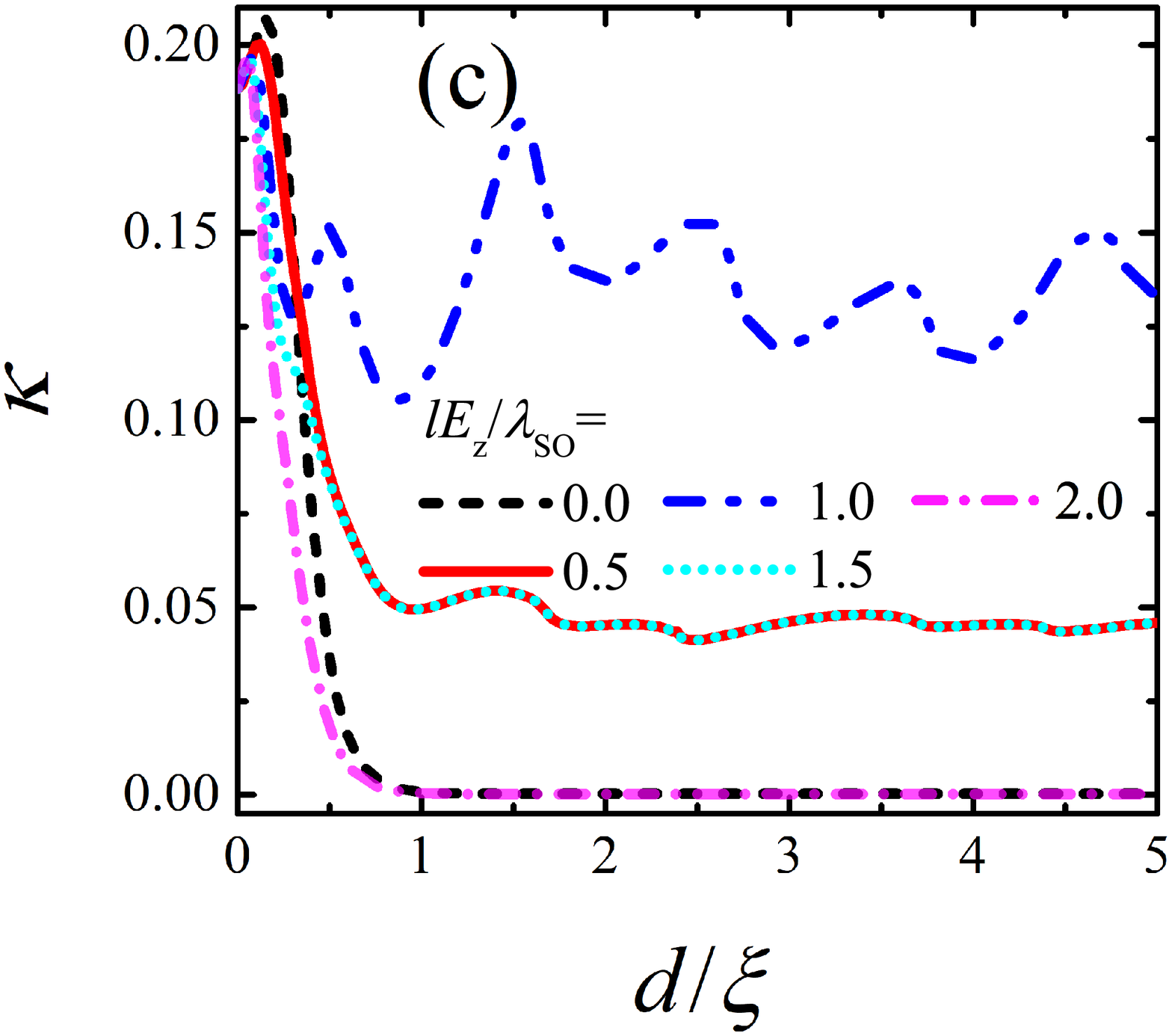}
  \label{subfig:fig3c}}
  \vspace{0.0cm}
  \hspace{2.0cm}
  \subfigure{
  \includegraphics[width=7cm]{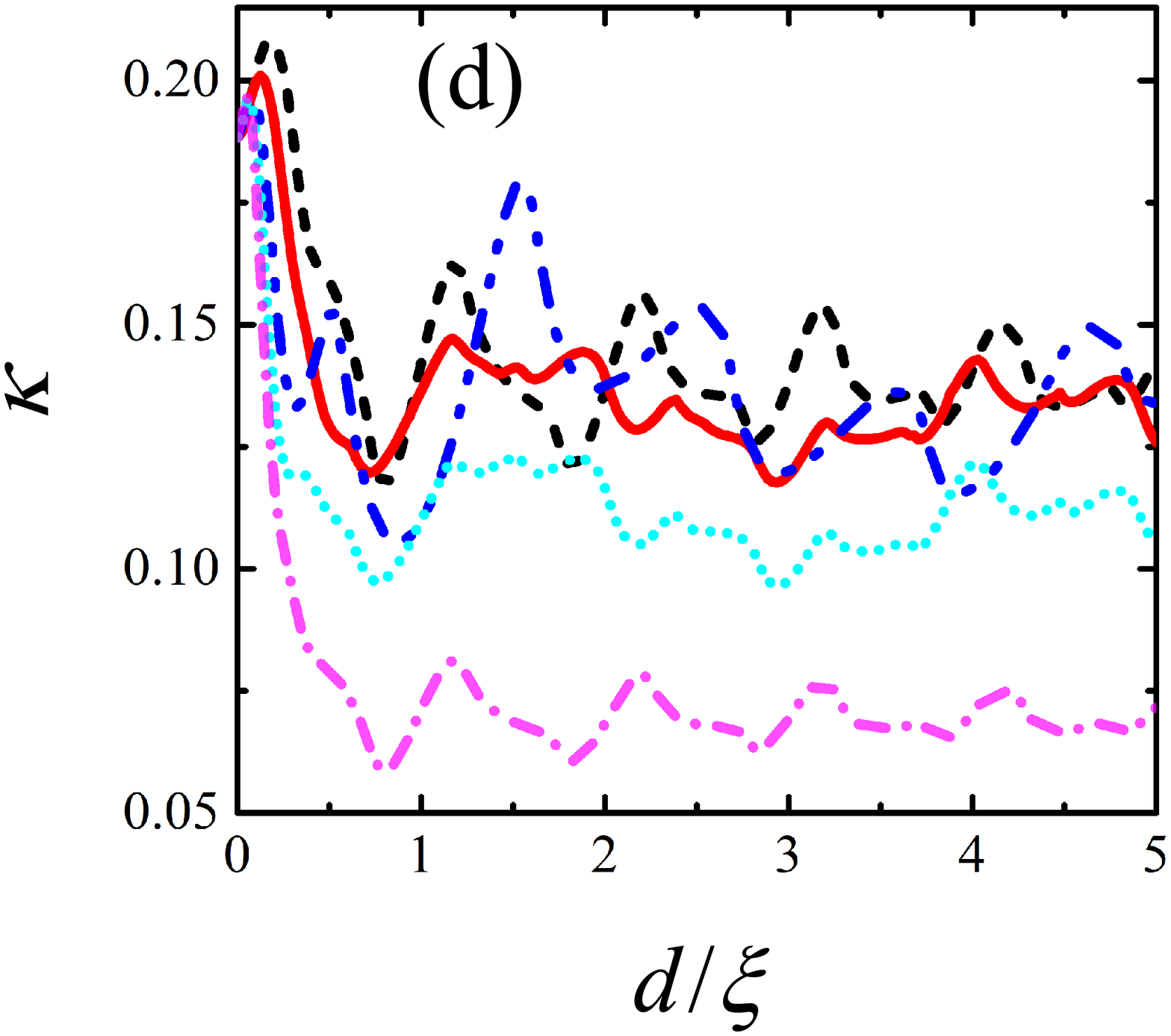}
  \label{subfig:fig3d}}
  \vspace{0.0cm}
  \hspace{0.0cm}
  \caption{(Color online) Normalized thermal conductance $\kappa$ as a function of the junction length $d$ with, where $\mu_M = 0$ in (a) and (c), and $\mu_M = \lambda_\mathrm{SO}$ in (b) and (d). Panels (a) and (b) and panels (c) and (d) present the scenarios in the S-AF-S junction with $h_{AF}=0.4 \lambda_\mathrm{SO}$ and S-F-S junction with $h_F= 0.4 \lambda_\mathrm{SO}$, respectively. In all panels, $\phi=\pi$ and $T=T_C/2$.}
  \label{fig:fig3}
  \end{figure*}

  As a starting point, we focus on the manifestations of the perpendicular electric field in the phase-coherent thermal conductance. Fig.~\ref{fig:fig2} presents the $\phi$-dependent thermal conductance with different perpendicular electric fields. As can be seen, in all cases the thermal conductance can be effectively modulated by the perpendicular electric field. In the S-AF-S junction, as depicted in Figs.~\ref{subfig:fig2a} and~\ref{subfig:fig2b}, by varying the strength of perpendicular electric field, the $\phi$-dependent thermal conductance exhibits transitions between minimal and maximal values at $\phi=\pi$. In the S-F-S junction, the perpendicular electric field changes the value of $\phi$ corresponding to the maximal and/or minimal thermal conductance, thus significantly tailoring the pattern of $\phi$-dependent thermal conductance, as shown in Figs.~\ref{subfig:fig2c} and~\ref{subfig:fig2d}. The electrical tunability of the phase-coherent thermal conductance results from the $lE_z$-dependent band structures of BTDMs. Resorting to Eq.~(\ref{eq:Hamiltonian}), in the M regions of S-AF-S and S-F-S junctions, the electron-like (hole-like) band edges can be, respectively, formulated as $E^{\pm, e(h)}_{\eta, \sigma} = \pm(\mp) |lE_z - \eta \sigma \lambda_{\mathrm {SO}} -(+) \sigma h_{AF}|-(+)\mu_M$ and $\varepsilon^{\pm, e(h)}_{\eta, \sigma}= \pm(\mp) |lE_z - \eta \sigma \lambda_{\mathrm {SO}}|- \sigma h_F -(+)\mu_M$, with the conduction (valence) band edge being indicated by the superscript $+(-)$ of $E^{\pm, e(h)}_{\eta, \sigma}$ and $\varepsilon^{\pm, e(h)}_{\eta, \sigma}$. Accordingly, in both junctions the perpendicular electric field can effectively tune the band edges which, in turn, regulate the band gaps of M regions. Since the quasiparticle transmission probabilities are profoundly influenced by the band gaps, the phase dependence of thermal conductance can be controlled by the perpendicular electric field. We note that the electrical tunability of the phase-coherent thermal conductance originates from the unique buckled sublattice structures of BTDMs and is absent in similar conventional \cite{Hajiloo2019, Pershoguba2019, Kazumi1965, Zhao2004, Zhao2003, Guttman1997} and topological Josephson junctions \cite{Sothmann2016, Li2017, Sothmann2017, Mukhopadhyay2021, Bauer2021L, Gresta2021, Mukhopadhyay2022, Bours2019, Blasi2020, Bours2018, Wang2022}. In addition, the configurations of $\phi$-dependent thermal conductance strongly depend on the type of exchange field and the chemical potential $\mu_M$. We illustrate the underlying physics resorting to the band structures of M regions.

  \begin{figure*}
  \centering
  \subfigure{
  \includegraphics[width=7cm]{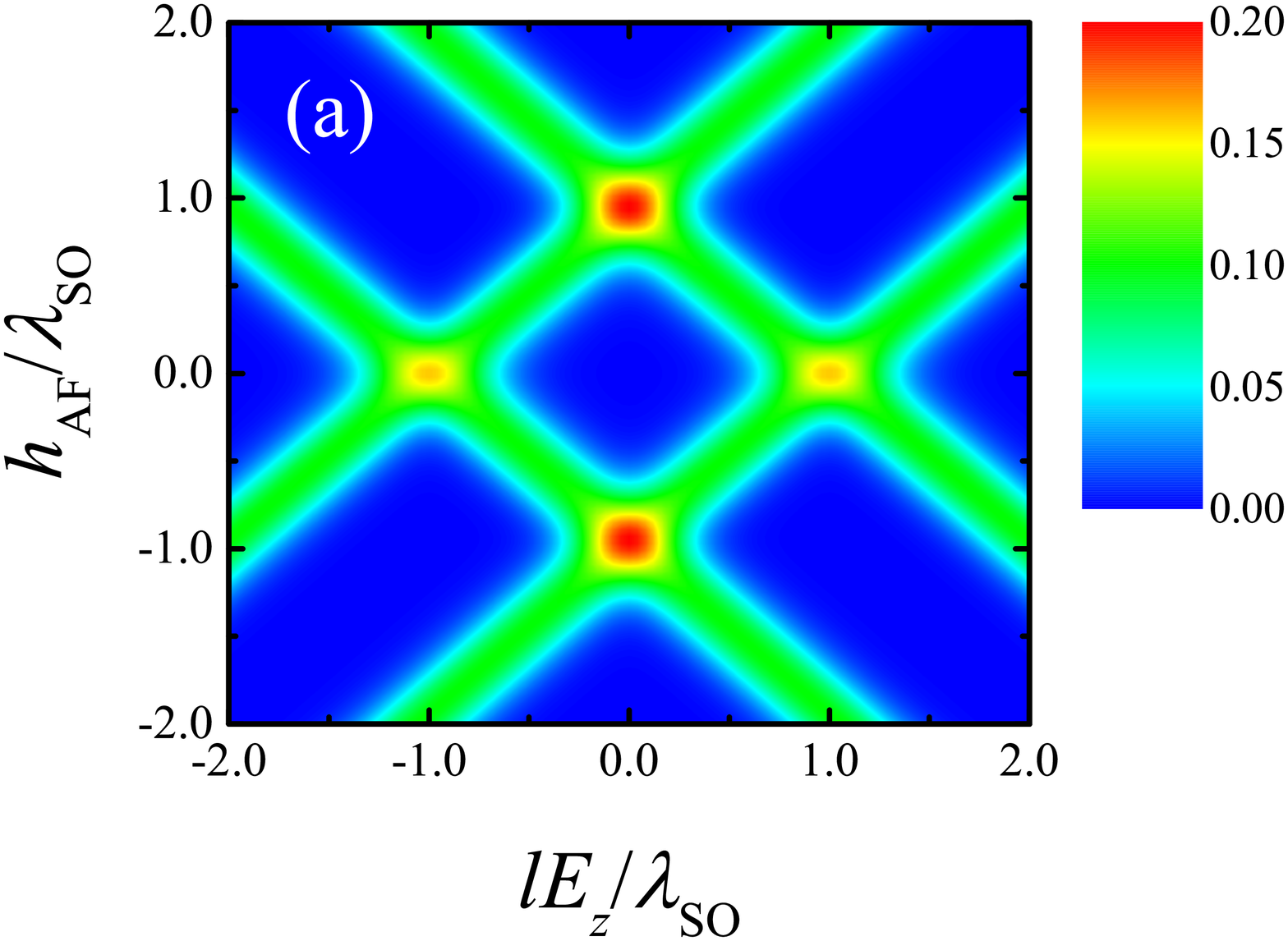}
  \label{subfig:fig4a}}
  \vspace{0.0cm}
  \hspace{2.0cm}
  \subfigure{
  \includegraphics[width=7cm]{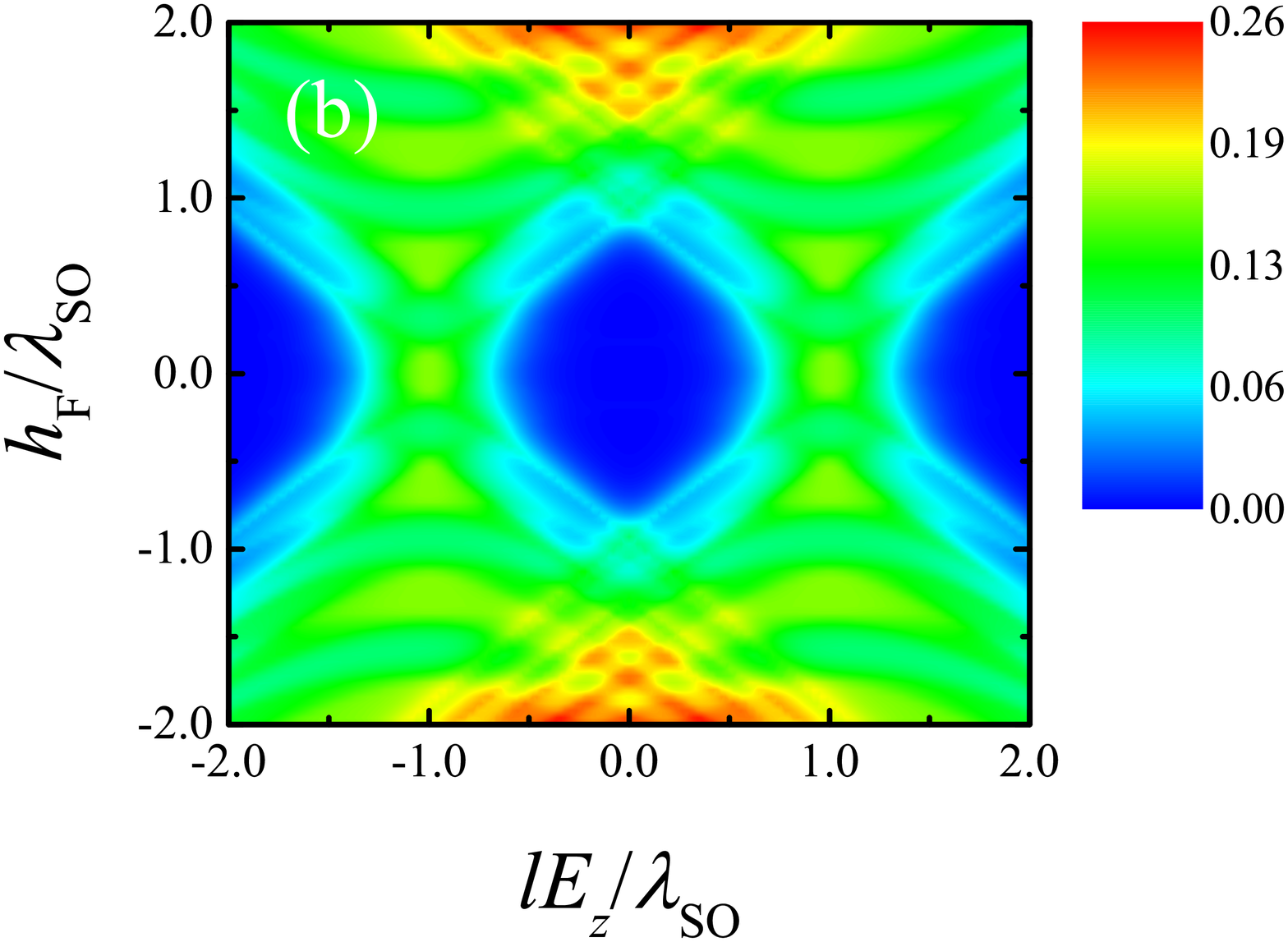}
  \label{subfig:fig4b}}
  \vspace{0.0cm}
  \hspace{0.0cm}
  \caption{(Color online) Contour plots of the normalized thermal conductance $\kappa$ of (a) S-AF-S junction and (b) S-F-S junction, where $\phi=\pi$, $d=\xi$, $\mu_M=0$, and $T=T_C/2$. }
  \label{fig:fig4}
  \end{figure*}

  \par
  Since the thermal conductance is essentially contributed by the propagating quasiparticles with energies just above the superconducting gap \cite{Kazumi1965, Guttman1997, Zhao2003, Zhao2004, Sothmann2016}, in the M region of S-AF-S (S-F-S) only the branches with band edges satisfying $E^{+(-), e(h)}_{\eta, \sigma} \le \Delta(T)$ ($\varepsilon^{+(-), e(h)}_{\eta, \sigma} \le \Delta(T)$) or $E^{-(+), e(h)}_{\eta, \sigma} \ge -\Delta(T)$ ($\varepsilon^{-(+), e(h)}_{\eta, \sigma} \ge -\Delta(T)$) can dominate the thermal transport. Keeping this principle in mind, we first discuss the scenarios in the S-AF-S junction with a undoped M region, i.e., $\mu_M=0$. According to the parameters fixed in Figs.~\ref{fig:fig2} and~\ref{fig:fig3}, in the situations of $lE_z=0$, $\lambda_{\mathrm {SO}}$, and $2 \lambda_{\mathrm {SO}}$, all band edges are outside of the regime of $[-\Delta(T), \Delta(T)] \simeq [-0.2\lambda_{\mathrm {SO}}, 0.2\lambda_{\mathrm {SO}}]$, so that the thermal transport is determined by the evanescent modes. Consequently, as shown in Fig.~\ref{subfig:fig3a}, the thermal conductance exponentially decays by increasing the junction length for $lE_z=0$, $\lambda_{\mathrm {SO}}$, and $2 \lambda_{\mathrm {SO}}$. When the junction length $d=\xi$ being large enough (c.f. Fig.~\ref{subfig:fig3a}), the thermal conductance contributed by the evanescent modes is strongly suppressed in the cases of $lE_z=0$, $\lambda_{\mathrm {SO}}$, and $2 \lambda_{\mathrm {SO}}$, as depicted in Fig.~\ref{subfig:fig2a}. While for $lE_z=0.5 \lambda_{\mathrm {SO}}$ and $1.5 \lambda_{\mathrm {SO}}$, there are two branches with band edges being located in the regime of $[-\Delta(T), \Delta(T)]$, i.e., $E^{\pm, e}_{-1, -1}|_{lE_z=0.5 \lambda_{\mathrm {SO}}}=E^{\mp, h}_{+1, +1}|_{lE_z=0.5 \lambda_{\mathrm {SO}}}= \pm 0.1 \lambda_{\mathrm {SO}}$ and $E^{\pm, e}_{+1, +1}|_{lE_z=1.5 \lambda_{\mathrm {SO}}}=E^{\mp, h}_{-1, -1}|_{lE_z=1.5 \lambda_{\mathrm {SO}}}= \pm 0.1 \lambda_{\mathrm {SO}}$, respectively. Therefore, in the situation of $lE_z=0.5 \lambda_{\mathrm {SO}}$ ($1.5 \lambda_{\mathrm {SO}}$), the thermal conductance is mainly contributed by the spin-down (spin-up) electron-like propagating quasiparticles of $K^\prime$ ($K$) valley and the spin-up (spin-down) hole-like propagating quasiparticles stemming from $K$ ($K^\prime$) valley, as confirmed by the oscillating profile in the $d$-dependent thermal conductance shown in Fig.~\ref{subfig:fig3a}. When the M region is lightly doped with $\mu_M=\lambda_{\mathrm {SO}}$, for each value of $lE_z$ selected in Figs.~\ref{fig:fig2} and~\ref{fig:fig3}, there exist at least two branches can approach in the regime of $[-\Delta(T), \Delta(T)]$ to support propagating quasiparticles. Taking the scenario of $lE_z=2\lambda_{\mathrm {SO}}$ as an example, in this case the band edges $E^{+, e}_{+1, +1}|_{lE_z=2\lambda_{\mathrm {SO}}}=-0.4\lambda_{\mathrm {SO}}<\Delta(T)$ and $E^{+, h}_{-1, -1}|_{lE_z=2\lambda_{\mathrm {SO}}}=0.4\lambda_{\mathrm {SO}}>-\Delta(T)$, thus both the spin-up electron-like branch of $K$ valley and the spin-down hole-like branch of $K^\prime$ valley can support propagating quasiparticles. As shown in Fig.~\ref{subfig:fig3b}, the thermal conductance exhibits pronounced oscillating profiles with respect to the junction length.

  \par
  In the S-F-S junction, the dependence of the phase-coherent thermal conductance on $lE_z$ can be understood in the similar way mentioned above. Here we only concentrate on the distinct scenarios that are absent in the S-AF-S junction. In the S-F-S junction, although the band edges of M region rely on the ferromagnetic exchange field $h_F$, the band gap scales $\delta^{F, e}_{\eta, \sigma}=\delta^{F, h}_{\eta, \sigma}=2|lE_z - \eta \sigma \lambda_{\mathrm {SO}}|$ are independent of $h_F$, in contrast to the manifestation of antiferromagnetic exchange field in the S-AF-S junction. Therefore, for a set of fixed valley and spin indices, the band gap in the M region of S-F-S junction is solely determined by the perpendicular electric field. This character leads to intriguing consequences in the special case of $lE_z=\lambda_{\mathrm {SO}}$, where the branches with $(\eta, \sigma)=(+1, +1)$ and $(\eta, \sigma)=(-1, -1)$ are gapless, thus the spin-up quasiparticles originating from $K$ valley and the spin-down quasiparticles stemming from $K^\prime$ valley can always contribute to the thermal transport. Consequently, when $lE_z=\lambda_{\mathrm {SO}}$ the thermal conductance obviously oscillates with $d$, regardless of the values of $\mu_M$, as shown in Figs.~\ref{subfig:fig3c} and~\ref{subfig:fig3d}.

  \par
  To further elucidate the different manifestations of antiferromagnetic and ferromagnetic exchange fields in the thermal transport, in Fig.~\ref{fig:fig4} we present the contour plots of the normalized thermal conductance in the $(lE_z, h)$ plane with $\mu_M=0$. In the S-AF-S junction, the antiferromagnetic exchange field not only changes the band edges, but also regulates the energy gap as $\delta^{AF, e(h)}_{\eta, \sigma}=2|lE_z - \eta \sigma \lambda_{\mathrm {SO}}-(+)\sigma h_{AF}|$. Since the band gap of M region depends both on the antiferromagnetic exchange field and on the perpendicular electric field, in the $(lE_z, h_{AF})$ plane the non-vanishing thermal conductance can only appear in the regions restricted by the conditions of $\pm h_{AF} + 0.8\lambda_{\mathrm{SO}} \le lE_z \le \pm h_{AF} + 1.2\lambda_{\mathrm{SO}}$ and $\pm h_{AF} -1.2 \lambda_{\mathrm{SO}} \le lE_z \le \pm h_{AF} - 0.8\lambda_{\mathrm{SO}}$, as shown in Fig.~\ref{subfig:fig4a}. However, in the S-F-S junction the ferromagnetic exchange field only shifts the position of band gap, but does not change its scale. For a set of fixed spin and valley indices, the band gaps of electron-like and hole-like branches share the same value, i.e, $\delta^{F, e}_{\eta, \sigma}=\delta^{F, h}_{\eta, \sigma}=2|lE_z - \eta \sigma \lambda_{\mathrm {SO}}|$. Therefore, in the case of $lE_z = +(-)\lambda_{\mathrm{SO}}$ both the electron-like and hole-like branches with $\eta\sigma=+(-)1$ are gapless to support propagating quasiparticles, regardless of the value of $h_F$. As presented by Fig.~\ref{subfig:fig4b}, in the whole range of $h_F$ the thermal conductance keeps finite at $lE_z = \pm \lambda_{\mathrm{SO}}$, this character is quite different from that in the S-AF-S junction. Furthermore, since the ferromagnetic exchange field can shift the positions of band gaps without affecting their scales, when the ferromagnetic field is large enough the band gaps can be pushed outside of the regime of $[-\Delta(T), \Delta(T)]$. As a consequence, the thermal conductance is non-vanishing when $|h_F|> 1.5\lambda_{\mathrm {SO}}$, this phenomenon is also distinct from that in the S-AF-S junction.

\section{\label{sec:level4} Conclusion}

  In conclusion, we have theoretically studied the thermal transport properties in BTDM-based S-AF-S and S-F-S junctions by virtue of the scattering wave approach. We have revealed that, in both S-AF-S and S-F-S junctions, the phase dependence of thermal conductance can be effectively controlled by perpendicular electric fields. This scenario results from the exotic buckled sublattice geometries of BTDMs and is absent in similar conventional and topological Josephson junctions. Resorting to the band structures in the M regions, we have illustrated the underlying mechanism behind the electrical tunability of thermal conductance. The different influences of the antiferromagnetic and ferromagnetic exchange fields on the thermal conductance have also been elucidated in detail. These results suggest that the BTDM-based Josephson junctions provide unique platforms for obtaining electrically tunable phase-coherent thermal transport, and we anticipate more interesting results for the thermal transport properties regarding the crossed Andreev reflections in BTDM-based superconducting hybrid structures.

\begin{acknowledgments}
 We acknowledge helpful discussions with R. Wang. This work was supported by the Science and Technology Planning Project of Hunan Province (Grant No. 2019RS2033), the National Natural Science Foundation of China (Grants No. 11804091 and  No. U2001215), the Hunan Provincial Natural Science Foundation (Grant No. 2019JJ50380), and the excellent youth fund of the Hunan Provincial Education Department (Grant No. 18B014).
\end{acknowledgments}

%\vbox{}

\appendix

\section{\label{sec:levela}Calculation of the basis scattering states in BTDM-based S-AF-S and S-F-S junctions}

  In this Appendix we give necessary calculation details regarding the wave functions and related parameters in the BTDM-based S-AF-S and S-F-S junctions.

  \par
  We assume that the translational symmetry in the proposed setup is preserved in the $y$ direction, so that the transverse momentum $k_y$ can be treated as a good quantum number. Under this assumption, in the S regions solving the BdG equation ${\cal H} (-i\partial_x, k_y) \psi = \epsilon \psi$ straightforwardly yields
  \begin{subequations}
  \begin{equation}
  \psi^{{L(R)}, \pm}_{eq} =
  \left( {\begin{array} {*{20}c}
  \pm \eta \sigma e^{\pm i \eta \theta^{L(R)}_{eq}} u^{L(R)} \\
  \sigma \gamma^{L(R)}_{eq} u^{L(R)} \\
  \pm \eta e^{\pm i \eta \theta^{L(R)}_{eq} - i\phi_{L(R)}} v^{L(R)} \\
  \gamma^{L(R)}_{eq} e^{-i\phi_{L(R)}} v^{L(R)} \\
  \end{array}} \right) e^{\pm i k^{L(R)}_{eq} \cos \theta^{L(R)}_{eq} x},
  \label{eq:WFSeq}
  \end{equation}
  \begin{equation}
  \psi^{{L(R)}, \pm}_{hq} =
  \left( {\begin{array} {*{20}c}
  \mp \eta \sigma e^{\mp i \eta \theta^{L(R)}_{hq}} v^{L(R)} \\
  \sigma \gamma^{L(R)}_{hq} v^{L(R)} \\
  \mp \eta e^{\mp i \eta \theta^{L(R)}_{hq} - i\phi_{L(R)}} u^{L(R)} \\
  \gamma^{L(R)}_{hq} e^{-i\phi_{L(R)}} u^{L(R)} \\
  \end{array}} \right) e^{\mp i k^{L(R)}_{hq} \cos \theta^{L(R)}_{hq} x },
  \label{eq:WFShq}
  \end{equation}
  \label{eq:WFS}
  \end{subequations}
  where we omit the trivial factor $e^{i k_y y}$, and the involved parameters take the forms of
  \begin{subequations}
  \begin{equation}
  k^{L, R}_{eq(hq)}=\sqrt{\left(\mu_S +(-){\mathrm{sgn}} (\epsilon)\sqrt{\epsilon^2-\Delta^2_{L, R}(T_{L, R})}\right)^2-m^2_{\eta \sigma}}/(\hbar v_F),
  \end{equation}
  \begin{equation}
  \gamma^{L, R}_{eq(hq)}=\left(\sqrt{(\hbar v_F k^{L, R}_{eq(hq)})^2+m^2_{\eta \sigma}}-m_{\eta \sigma}\right)/\left(\hbar v_F k^{L, R}_{eq(hq)}\right),
  \end{equation}
  \begin{equation}
  \theta^{L, R}_{eq(hq)}=\sin^{-1}\left(k_y/k^{L, R}_{eq(hq)}\right),
  \end{equation}
  \begin{equation}
  u^{L, R}=\sqrt{\frac{1}{2}\left(1+\sqrt{1-\Delta^2_{L, R}(T_{L, R})/\epsilon^2}\right)},
  \end{equation}
  \begin{equation}
  v^{L, R}={\mathrm{sgn}} (\epsilon) \sqrt{\frac{1}{2}\left(1-\sqrt{1-\Delta^2_{L, R}(T_{L, R})/\epsilon^2}\right)}.
  \end{equation}
  \end{subequations}

  \par
  In the magnetic region ($0<x<d$), after omitting the trivial factor $e^{i k_y y}$, the scattering states can be formulated as
  \begin{subequations}
  \begin{equation}
  \psi^\pm_e =
  \left( {\begin{array} {*{20}c}
  \pm \eta s_e e^{\pm i \eta s_e \alpha_e} \\
  \gamma_e \\
  0 \\
  0 \\
  \end{array}} \right) e^{\pm i s_e k_e \cos \alpha_e x},
  \label{eq:WFMe}
  \end{equation}
  \begin{equation}
  \psi^\pm_h =
  \left( {\begin{array} {*{20}c}
  0 \\
  0 \\
  \mp \eta s_h e^{\pm i \eta s_h \alpha_h}  \\
  \gamma_h \\
  \end{array}} \right) e^{\pm i s_h k_h \cos \alpha_h x},
  \label{eq:WFMh}
  \end{equation}
  \label{eq:WFM}
  \end{subequations}
  where the scattering angle $\alpha_{e(h)}=\sin^{-1}\left( k_y/k_{e(h)} \right)$.

  \par
  In the S-AF-S junction, the related parameters in Eq.~(\ref{eq:WFM}) are given by
  \begin{subequations}
  \begin{equation}
  k_{e(h)}=\sqrt{(\epsilon +(-) \mu_M)^2-(m_{\eta\sigma}-(+)\sigma h_{AF})^2}/(\hbar v_F),
  \end{equation}
  \begin{equation}
  \gamma_{e(h)}=\left(\epsilon +(-) \mu_M -(+) m_{\eta \sigma} + \sigma h_{AF}\right)/(\hbar v_F k_{e(h)}),
  \end{equation}
  \begin{equation}
   s_{e(h)}={\mathrm{sgn}}\left(\epsilon - |m_{\eta\sigma}-(+)\sigma h_{AF}|+(-)\mu_M\right).
  \end{equation}
  \end{subequations}
  While in the S-F-S junction, the corresponding parameters in Eq.~(\ref{eq:WFM}) are defined as
  \begin{subequations}
  \begin{equation}
  k_{e(h)}=\sqrt{(\epsilon +(-) \mu_M + \sigma h_F)^2-m^2_{\eta\sigma}}/(\hbar v_F),
  \end{equation}
  \begin{equation}
  \gamma_{e(h)}=\left[\epsilon +(-) \mu_M -(+) m_{\eta \sigma} + \sigma h_F \right]/(\hbar v_F k_{e(h)}),
  \end{equation}
  \begin{equation}
   s_{e(h)}={\mathrm{sgn}}\left(\epsilon - |m_{\eta\sigma}| +(-) \mu_M + \sigma h_F \right).
  \end{equation}
  \end{subequations}

\vbox{}
%\nocite{*}

%\bibliography{bibTex}


\end{document}